\documentclass[
    ,final            
  ]
  {aipproc}

\usepackage{bbm}
\usepackage{amsmath}
\usepackage{amssymb}

\layoutstyle{8x11single}

\newcommand{\be}{\begin{equation}}
\newcommand{\ee}{\end{equation}}
\newcommand{\eea}{\end{eqnarray}}
\newcommand{\bea}{\begin{eqnarray}}

\newcommand{\exs}[1]{\ensuremath{\langle{#1}\rangle}}

\newcommand{\eins}{\ensuremath{\mathbbm 1}}

\newcommand{\WW}{\ensuremath{\mathcal{W}}}

\newcommand{\ketbra}[1]{\ensuremath{| #1 \rangle \langle #1 |}}

\newcommand{\ket}[1]{\ensuremath{|#1\rangle}}

\newcommand{\kommentar}[1]{}

\begin{document}

\title{Two measurement settings can suffice to 
verify multipartite entanglement }


\author{G\'eza T\'oth}{
  address={Theoretical Division, Max Planck Institute 
for Quantum Optics, Hans-Kopfermann-Stra{\ss}e 1,\\ 
D-85748 Garching, Germany},
  homepage={http://www.},
  email={Geza.Toth@mpq.mpg.de},
  thanks={}
}

\author{Otfried G\"uhne}{
  address={Institut f\"ur Quantenoptik und Quanteninformation,
\"Osterreichische Akademie der Wissenschaften,\\
A-6020 Innsbruck, Austria},
  altaddress={Institut f\"ur Theoretische Physik, Universit\"at
Hannover, Appelstra{\ss}e 2, D-30167 Hannover, Germany}
}

\copyrightyear{2001}

\begin{abstract}
We present entanglement witnesses for detecting {\it genuine} 
multi-qubit entanglement. Our constructions are robust against
noise and require only two {\it local} measurement 
settings, independent of the number of qubits. Thus  they 
allow to verify entanglement of many qubits in experiments 
while requiring only a small effort. 
In contrast, usual methods 
need an effort which increases exponentially 
with the number of qubits.
The witnesses detect states close to 
GHZ states and cluster states.

\end{abstract}

\date{\today}

\maketitle

\section{Introduction}

Entanglement, a strange phenomenon of the quantum world, 
has been known since the first half of the previous 
century. Recently, new insight on entanglement was gained 
through quantum information science  which 
connects physics with algorithmic theory. In this context, 
besides asking "What are the characteristics of an entangled 
state?", we can also ask "What kind of tasks can be done with 
entangled states?" or "What types of  entangled states 
can be created?"

Questions of the second type lead to the classification 
of multi-partite entangled states. For two-qubits it is 
enough to say: "This state is entangled" or "This state 
is separable". From an ensemble of two-qubit systems it 
is always possible to distill with local operations and 
classical communication (LOCC) a maximally entangled singlet 
state, if the corresponding density matrix is entangled. 

But already for three qubits, the situation is much more 
complicated. First of all, we have to differentiate the 
case when two qubits are entangled and the third is not 
entangled with them (e.g., $\ket{\phi_1}=
\ket{0}(\ket{00}+\ket{11})/\sqrt{2}$) 
from real three qubit entanglement 
(e.g., $\ket{\phi_2}=(\ket{000}+\ket{111})/\sqrt{2}$). 
Moreover, given two genuine tripartite entangled three
qubit states, one may ask whether it is possible to convert 
a state into another one using only LOCC.
Surprisingly, it turns out that not all pure states with 
{\it genuine} three-qubit entanglement  are equivalent 
under LOCC, not even stochastically. In fact, there are 
two inequivalent classes, the W and the GHZ class
\cite{DV00}. This classification can be extended to
mixed states \cite{AB01} where the W class is inclusive 
of the GHZ class. For pure four qubit states, the number 
of equivalence classes is infinite \cite{FD02} and the 
extension of the classification to mixed states does not 
seem to be useful.

Thus for pure states of many qubits we are left with three 
qualitatively different cases: The {\it fully separable} states
are product states with no correlations between the parties.
For the {\it biseparable} states there always exists one partition
of the qubits into two parties, which are separable and not 
correlated. However, the qubits inside one party may be entangled.
For {\it genuine multipartite entangled} states no such splitting 
can be found.
A {\it mixed} state is biseparable (respectively, fully separable) if it 
can be constructed by mixing biseparable (respectively, fully separable) 
pure states. 

In this paper we will 
describe a method how to detect genuine $N$-qubit 
entanglement around GHZ (Greenberger-Horne-Zeilinger, 
\cite{GH90}) and cluster states \cite{BR03}.
Besides being theoretically interesting, the motivation for
detecting multi-qubit entanglement also comes from the side 
of the experimentalists. Recently, several experiments succeeded 
in creating various multi-qubit states with photons 
\cite{PB00}, trapped ions \cite{SK00} or cold atoms 
trapped with optical lattices \cite{MG03B}. For all these 
experiments it is a crucial to prove that the quantum state 
is genuine multipartite entangled: A multi-qubit experiment 
presents something qualitatively new only 
if provably more than two qubits are entangled.

\begin{figure}
\resizebox{3.5in}{!}{\includegraphics{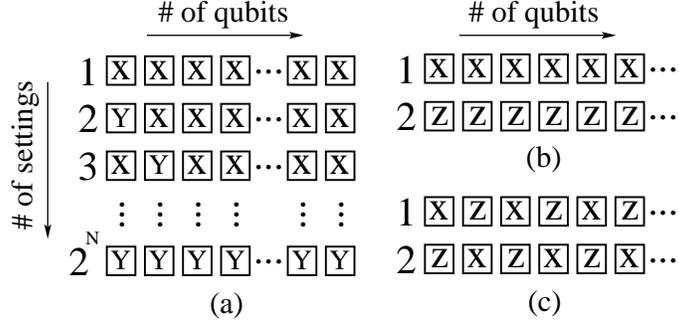}}
\caption{(a) The $2^N$ measurement settings needed 
using Bell inequalities for detecting
multi-qubit entanglement close to a GHZ state.  
For each qubit the measured spin component is indicated. 
(b) The two
settings needed for our witnesses detecting genuine multi-qubit
entanglement close to GHZ and (c) cluster states.} \label{FigGHZ}
\end{figure}

Detecting genuine multi-qubit entanglement is a difficult 
problem since it is inherently {\it nonlocal}, while in most 
experiments only {\it local} measurements are possible. One 
option is using Bell inequalities. These indicate the violation 
of local realism, a notion independent of quantum physics. 
However, in some cases they can be used not only for detecting 
quantum entanglement, but for detecting genuine $N$-qubit 
entanglement \cite{Bell}. Bell inequalities typically 
require measuring two variables at each qubit and computing 
an expression constructed as a sum of some $N$-qubit 
correlations. If the value of this expression is larger than 
a certain bound the system is $N$-qubit entangled.
The drawback of this method is that it 
requires using very many {\it measurement settings}. 
As shown in Fig.~\ref{FigGHZ}(a), they need typically 
$2^N$ local measurement settings for detecting entanglement 
close to GHZ states.

Let us shortly explain what we understand by such a local
measurement setting. Measuring a local setting 
$\{O^{(k)}\}$ with ${k=1,...,N}$ consists of simultaneously performing 
the von Neumann measurements $O^{(k)}$ 
on the corresponding qubits, indexed by $k.$ After repeating
the measurements several times, the coincidence 
probabilities for the outcomes are collected. 
For $N$ qubits there are $2^N$ different outcome probabilities. 
Given
these probabilities it is possible to compute all the two point 
correlations $\exs{O^{(k)}O^{(l)}}$, three-point correlations 
$\exs{O^{(k)} O^{(l)}O^{(m)}}$, etc. Since  all these 
correlation terms can be measured with one setting, 
the number of settings determines the experimental 
effort rather than the number of measured correlation
terms. Obviously, with one local measurement setting it is not
possible to detect entanglement. Thus, two measurement settings
are the minimal effort needed for the detection of entanglement.

In this paper we explain some of the ideas of Ref.~\cite{TG04} 
on the detection of genuine multi-qubit entanglement.
We will present entanglement conditions which require
only {\it two} settings independent from the number of qubits. 
Since the number of measurement settings needed for existing 
methods increases exponentially with the number of qubits, 
our conditions provide a very effective way to detect 
entanglement.
For instance, they need $2^{N-1}$ times less measurement settings 
than Bell inequalities. For a large number of qubits this 
difference is crucial --- the new conditions do not 
improve things only quantitatively, they make detection possible 
when it would be unrealistic otherwise.

Our conditions will be presented in the form of 
{\it entanglement witnesses}
\cite{HH90}. These are operators which have a positive or zero
expectation value for all separable states. Thus a negative
expectation value signals the presence of entanglement.
In constructing the entanglement witnesses for cluster states and
GHZ states we use the {\it stabilizing operators} of these states.
An observable $S$ is a stabilizing operator of a state $\ket{\Psi}$ 
if it satisfies 
\be
S \ket{\Psi} = \ket{\Psi}. 
\label{stabil} 
\ee 
In our case the operators $S$ are tensor products of Pauli spin 
matrices. They can, therefore, easily be measured locally.

\section{Detecting GHZ states}

Let us start with GHZ states. An $N$ qubit GHZ state is defined
as 
\be
\ket{GHZ_N}=\frac{1}{\sqrt{2}}(\ket{0...0}+\ket{1...1}).
\ee
As maximally entangled multi-qubit states, GHZ states are 
intensively studied \cite{M90} and have been realized 
in numerous experiments \cite{PB00,SK00}. The stabilizing 
operators of an $N$ qubit GHZ state are 
\bea
S_1^{(GHZ_N)}&:=& \prod_{k=1}^N \sigma_x^{(k)},
\nonumber
\\
S_k^{(GHZ_N)}&:=&\sigma_z^{(k-1)} \sigma_z^{(k)}; \;\;
k\in\{2,3,..,N\}. 
\label{eigenGHZ} 
\eea 
In fact, one can easily calculate that for these observables
$S_k^{(GHZ_N)}\ket{GHZ_N}=\ket{GHZ_N}$ holds. Note that
not only $S_k^{(GHZ_N)}$ stabilize the GHZ state, but 
any products of these operators does it as well. These 
operators form a commutative group and the $S_k^{(GHZ_N)}$
are the generators of the group.

One entanglement witness detecting genuine 
$N$-qubit entanglement close to GHZ states 
is given by 
\be 
\WW_{GHZ_N} := 
3\cdot\eins-2\bigg[\frac{S_1^{(GHZ_N)}+\eins}{2}
+\prod_{k=2}^N\frac{S_k^{(GHZ_N)}+\eins}{2}\bigg].
\label{wnoisetol} 
\ee 
The witness $\WW_{GHZ_N}$ uses only two measurement 
settings, namely the ones in Fig. 1(b). The structure 
of $\WW_{GHZ_N}$ given in Eq.~(\ref{wnoisetol}) can
be interpreted as follows. The two terms in the square 
brackets are two projectors. The first is a projector 
on the subspace for which $\exs{S_1^{(GHZ_N)}}=+1$. The 
second one is a projector on the subspace for which 
$\exs{S_k^{(GHZ_N)}}=+1$ for any $k\in\{2,3,...,N\}$.  
The GHZ state is the only state which is in both spaces, 
thus the mean value of 
$\WW_{GHZ_N}$ is $-1$ only for this state. For any other state it is larger.
In general, the more negative $\exs{\WW_{GHZ_N}}_{\ket{\Psi}}$ is,
the closer $\ket{\Psi}$ is, in some sense,  to the GHZ state
\cite{DISTANCE}. It is known that in the proximity of the GHZ state
there are only states with genuine $N$-qubit entanglement,
so the constant in Eq.~(\ref{wnoisetol}) is chosen such 
that if $\exs{\WW_{GHZ_N}}<0$ then the state is in this 
neighborhood and is detected as entangled.

From the practical point of view it is very important 
to know, how large the neighborhood of the GHZ state is which is
detected as entangled by the witness. This is usually 
characterized by the  robustness to noise. Let us 
consider a GHZ state mixed with white noise 
\be
\varrho(p_{noise}):=
p_{noise} \cdot \frac{\eins}{2^N} +
(1-p_{noise}) \ketbra{GHZ_N}.
\ee 
The witness $\WW_{GHZ_N}$ detects the state as entangled 
if $p_{noise}<1(3-4/2^N)$. The bound on noise is explicitly 
shown in Table I. Our witness is quite robust ---  it 
tolerates at least $33\%$ noise even for large $N$.

\section{Detecting cluster states}

Now let us turn to cluster states. These have recently 
raised a lot of interest both theoretically and 
experimentally. They can easily be created in a spin 
chain with Ising-type interaction \cite{BR03} and have 
been realized in optical lattices of two-state atoms 
\cite{MG03B}. Remarkably, their entanglement is more 
persistent to noise than that of a GHZ state \cite{BR03}.
They play a central role in error correction \cite{G96},
fault-tolerant quantum computation, cryptographic protocols such
as secret sharing \cite{CG99}, and measurement-based quantum
computation \cite{RB03}.

For three qubits the cluster state $\ket{C_3}$ is 
equivalent to a GHZ state up to local unitary 
transformations.
For four qubits the state $\ket{C_4}$ can be  transformed by 
some local unitaries into $\ket{\phi}=
(\ket{0000}+\ket{0011}+\ket{1100}-\ket{1111})/2.$
For an arbitrary number of qubits it is more 
convenient to use a general definition via 
stabilizing operators than writing it out explicitly in some basis.
The stabilizing operators of an $N$-qubit cluster state are 
\bea
S_1^{(C_N)}&:=&\sigma_x^{(1)} \sigma_z^{(2)},
\nonumber\\
S_k^{(C_N)}&:=&\sigma_z^{(k-1)} \sigma_x^{(k)} \sigma_z^{(k+1)};
k\in\{2,3,..,N-1\},
\nonumber\\
S_N^{(C_N)}&:=&\sigma_z^{(N-1)} \sigma_x^{(N)}. 
\label{eigenC}
\eea 
Given these stabilizing operators, the cluster state $\ket{C_N}$
is {\it defined} as the state fulfilling
\be
S_i^{(C_N)}\ket{C_N}=\ket{C_N}.
\ee
On can show that the cluster state is uniquely defined by 
these equations.  
Our witness for the detection of $N$-qubit entanglement
around a cluster state is 
\bea 
\WW_{C_N} &:=& 
3\cdot\eins - 
2\bigg[\prod_{\text{even k}}\frac{S_k^{(C_N)}+\eins}{2}+
\prod_{\text{odd k}}\frac{S_k^{(C_N)}+\eins}{2}\bigg]. 
\label{CN}
\eea 
If the expectation value of $\WW_{C_N}$ is negative then the
system is genuine $N$-qubit entangled. Again, only two settings 
are needed. These are shown in Fig. \ref{FigGHZ}(c). The witness
tolerates at least $25\%$ noise as shown in Table I. The structure of
$\WW_{C_N}$ is similar to that of $\WW_{GHZ_N}$. In the square
brackets there are two terms. The first term is a projector on the
subspace for which $\exs{S_k^{(C_N)}}=+1$ for even $k$. The second
term is a projector on the subspace for which
$\exs{S_k^{(C_N)}}=+1$ for odd $k$.

\section{Summary}

In summary, we have presented entanglement witnesses for detecting
genuine multi-qubit entanglement close to GHZ and cluster states.
These witnesses are easy to measure since they require only two
measurement settings. For further details, especially for the proofs of
the theorems presented here, please see
Ref. \cite{TG04}. This reference also describes how to generalize
the results for graph states.

\begin{table}
\caption{Noise tolerance for the GHZ state and the cluster state
witnesses vs. number of qubits}
\begin{tabular}{l|l|l|l|l|l|l|l|l|l|l}
N   &  2    & 3    & 4    & 5  & 6   & 7  & 8  & 9   & 10 \\
\hline $\ket{GHZ_N}$
    &  0.50 & 0.40 & 0.36 &0.35 &0.34 &0.34 &0.34 & 0.33 &0.33\\
    $\ket{C_N}$
    &  0.50 & 0.40 & 0.33 &0.31 &0.29 &0.28 &0.27 & 0.26 &0.26\\
\end{tabular}
\end{table}

\begin{theacknowledgments}
We thank M.~Aspelmeyer, H.J.~Briegel, D.~Bru{\ss}, {\v
C}.~Brukner, J.I.~Cirac, T.~Cubitt, J.~Eisert,
J.J.~Garc\'{\i}a-Ripoll, P.~Hyllus, M.~Lewenstein, A.~Sanpera,
M.M.~Wolf, and  M.~\.Zukowski for useful discussions.
We also acknowledge the support of the DFG (Graduierten\-kolleg
282), the EU projects RESQ and QUPRODIS. G.T. thanks the support of
the European Union (Marie Curie Individual Grant No. MEIF-CT-2003-500183).
\end{theacknowledgments}

\bibliography{sample}

\begin{thebibliography}{99}
\bibitem{DISTANCE} To be more specific, $\exs{\WW^{(GHZ_N)}}_{\ket{\Psi}}$
can be used to find a lower bound on the overlap
$|\exs{GHZ_N|\Psi}|^2$.
\end{thebibliography}

\end{document}